\documentclass[aps,prl,superscriptaddress,amsmath,amssymb,reprint]{revtex4-2}

\usepackage{graphicx}
\usepackage{dcolumn}
\usepackage{bm}
\usepackage[T1]{fontenc}

\begin{document}


\title{Crack propagation by activated avalanches during creep and fatigue from elastic interface theory}


\author{Tero M\"{a}kinen}
\email{Corresponding author: tero.j.makinen@aalto.fi}
\affiliation{Department of Applied Physics, Aalto University, P.O. Box 15600, 00076 Aalto, Espoo, Finland}
\author{Lumi Tuokkola}
\affiliation{Department of Applied Physics, Aalto University, P.O. Box 15600, 00076 Aalto, Espoo, Finland}
\author{Joonas Lahikainen}
\affiliation{Department of Applied Physics, Aalto University, P.O. Box 15600, 00076 Aalto, Espoo, Finland}
\author{Ivan V. Lomakin}
\affiliation{Department of Applied Physics, Aalto University, P.O. Box 15600, 00076 Aalto, Espoo, Finland}
\author{Juha Koivisto}
\affiliation{Department of Applied Physics, Aalto University, P.O. Box 15600, 00076 Aalto, Espoo, Finland}
\author{Mikko J. Alava}
\affiliation{Department of Applied Physics, Aalto University, P.O. Box 15600, 00076 Aalto, Espoo, Finland}
\affiliation{NOMATEN Centre of Excellence, National Centre for Nuclear Research, A. Soltana 7, 05-400  Otwock-Świerk, Poland}


\date{\today}

\begin{abstract}
The growth of cracks combines materials science, fracture mechanics, and statistical physics. The importance of fluctuations in the crack velocity is fundamental since it signals that the crack overcomes local barriers such as tough spots by avalanches. In ductile materials the omnipresent plasticity close to the crack tip influences the growth by history effects, which we here study in 
polymethylmetacrylate by various fatigue and creep protocols. We show how the crack tip local history may be encompassed in a time- and protocol dependent lengthscale, that allows to apply a statistical fracture description to the time-dependent crack growth rate, resolving the well-known paradox why fatigue cracks grow faster if the stress during a cycle is let to relax more from the peak value. The results open up novel directions for understanding fracture by statistical physics.
\end{abstract}


\maketitle

Fracture is a multiscale problem involving microscopic material properties and continuum-level elastic fields. Time-dependent fracture can be categorized into creep---failure under constant load---and fatigue, failure under cyclic loading. These phenomena are commonly modeled using fracture mechanics combined with empirical and theoretical scaling laws, such as the Paris--Erdo\v{g}an law for fatigue crack growth \cite{ritchie1999mechanisms,ritchie2005incomplete}.

The statistical physics framework of subcritical depinning, originally developed for systems like magnetic domain walls~\cite{lemerle1998domain, paruch2006nanoscale, metaxas2007creep, kim2009interdimensional, jeudy2016universal, duttagupta2016adiabatic, pardo2017universal, jeudy2018pinning, albornoz2021universal}, superconducting vortices~\cite{anderson1964hard, troyanovski1999collective}, crumpled sheets~\cite{shohat2023logarithmic},
and elastic interfaces~\cite{chauve2000creep, muller2001velocity, kolton2009creep, wiese2022theory,purrello2017creep} offers insight into creep behavior. In this context, the velocity of a moving interface, such as a crack front, follows an Arrhenius law at small driving forces
\begin{equation} \label{eq:arrhenius}
    v = v_{\rm c} \exp\left( - \beta \Delta E \right)
\end{equation}
where $v_{\rm c}$ is the characteristic velocity, $\beta$ is the thermodynamic beta, and $\Delta E$ is the energy barrier. Due to the glassy nature of pinned interfaces, $\Delta E$ diverges as the driving force $G$ decreases ~\cite{feigel1989theory, blatter1994vortices}, following at small driving
\begin{equation} \label{eq:energyBarrier}
    \Delta E = \frac{1}{\beta_{\rm d}} \left[ \left( \frac{G}{G_{\rm c}} \right)^{-\mu} - 1 \right] . 
\end{equation}
Here, $\beta_{\rm d}$ represents the disorder energy scale, $G_{\rm c}$ is the critical driving force, and $\mu$, the barrier exponent~\cite{chauve2000creep, wiese2022theory}, depends on the material's elasticity, often taking values of $1/4$ for short-range elasticity and 1 for long-range elasticity in 1+1 dimensions.

\begin{figure}[th!]
    \centering
    \includegraphics[width=\columnwidth]{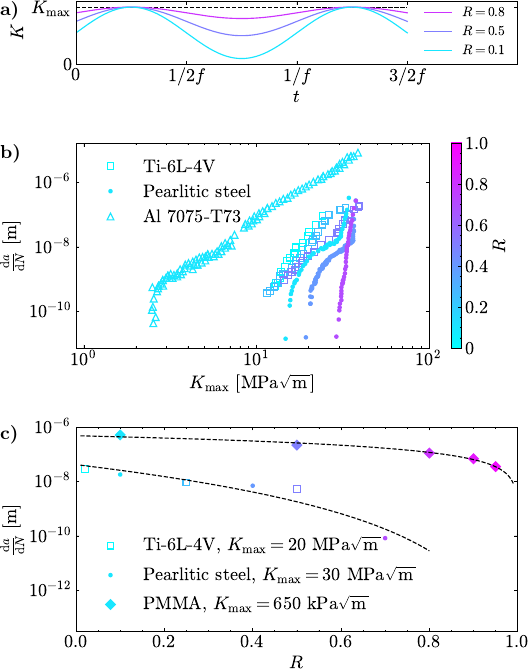}
    \caption{
    \textbf{a)}~The effect of the cycle contrast parameter $R$ on the SIF $K$ as a function of time $t$ when a constant $K_{\rm max}$ and frequency~$f$ are used. The dashed line corresponds to the creep limit $R \to 1$.
    \textbf{b)}~Typical Paris-curve plots of the SIF vs. crack velocity in example titanium~\cite{huang2007improved}, steel~\cite{el2004effects} and aluminum~\cite{forth2004anomolous} alloys with the cycle contrast parameter $R$ shown in color code: smaller $R$ leads to faster crack propagation.
    Transition from the Paris--Erdo\v{g}an regime (low $R$) to faster crack propagation (high $R$) is also seen.
    \textbf{c)}~The impact of~$R$ on velocity at a constant SIF for two metallic materials, as well as our PMMA experiments, showing the large decrease with $R$ in the slow crack regime. The dashed lines serve as a guide to the eye.
    }
    \label{fig:fig1}
\end{figure}

In fracture mechanics, similar principles apply ~\cite{koivisto2007creep,bonamy2008crackling, tallakstad2011local, Bares2014,Bares2018}. The energy release rate $G$, which drives crack growth, relates 
to the more familiar stress intensity factor (SIF) $K$ through $G = K^2 / E$, where $E$ is the material’s Young’s modulus. After an initial nucleation phase, crack growth during fatigue~(Fig.~1a) is described by the Paris--Erdo\v{g}an law~\cite{paris1963critical}, which relates the crack length increment $\mathrm{d} a$ per loading cycle $N$ to the SIF range $\Delta K = K_{\rm max} - K_{\rm min}$ as
\begin{equation} \label{eq:paris}
    \frac{\mathrm{d} a}{\mathrm{d} N} \propto \Delta K^m ,
\end{equation}
where $m$ is a material-dependent exponent. The Paris law shows apparent self-similarity~\cite{ritchie2005incomplete} but is typically valid only over a limited $\Delta K$ range~(Fig.~\ref{fig:fig1}b). Fatigue experiments~reveal a counterintuitive trend: reducing the minimum SIF $K_{\rm min}$ while maintaining $K_{\rm max}$ accelerates crack growth, an observation \cite{ZERBST2016190} sometimes summarized as “less stress makes cracks faster”~(Fig.~\ref{fig:fig1}c). This behavior highlights the significance of $\Delta K$ as the primary driver of fatigue growth, rather than the peak $K$.

Attempting to extend creep models to fatigue involves assuming that crack propagation arises from creep under a varying force. However, this approach struggles to explain why increasing $K_{\rm min}$ slows crack growth, contrary to creep-based expectations~\cite{savolainen2023velocity}. The main challenges in fatigue mechanics lie in accounting
 for the cyclic loading history and the role of $K_{\rm min}$ and how these affect the energy barrier (Eq.~\ref{eq:arrhenius}).
 Mechanisms like crack closure and effects specific to short cracks, modeled through elastic-plastic fracture mechanics (EPFM), have been invoked to address these issues~\cite{elber1970fatigue, pippan2017fatigue,zerbst2016fatigue,zerbst2018fatigue}. These insights are crucial for understanding how the energy barrier evolves during cyclic fatigue experiments and how it depends on parameters like $K_{\rm min}$.\\

\begin{figure}[th!]
    \centering
    \includegraphics[width=\columnwidth]{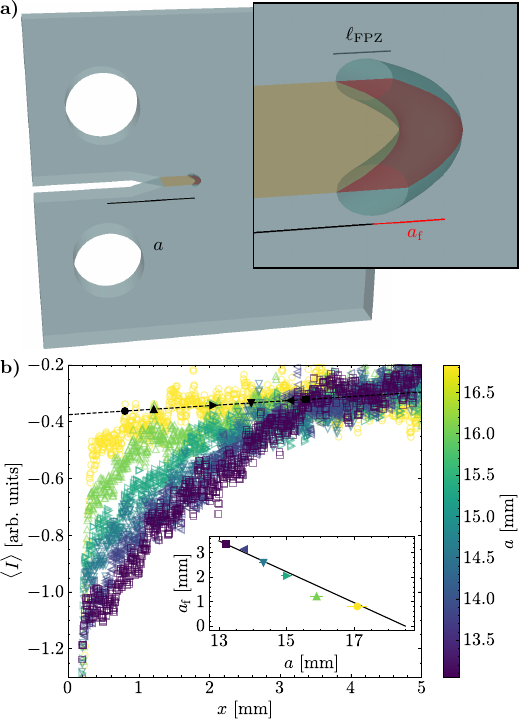}
    \caption{
    \textbf{a)}~The geometry of crack propagation in our experiment with a fracture process zone~(FPZ) of size $\ell_{\rm FPZ}$ to be found in front of the crack of size $a$ and the history effect, which is encapsulated in the lengthscale~$a_{\rm f}$ (see below). 
    \textbf{b)}~In PMMA one may measure a lengthscale in front of the crack by the optical variation of the material (see Methods and Ref.~\cite{supplementary} for details). 
    We show here average image intensity $\langle I \rangle$ in front of the crack for six different points during an example experiment with $R=0.1$, the color indicating the corresponding $a$-value.
    The point $x=0$~mm corresponds to the $a$-value for each curve, and the dashed line corresponds to the linear fit to the far-field intensity. The observed values of lengthscale~$a_{\rm f}$ are then the black markers on this line.
    The inset shows the observed values of the lengthscale~$a_{\rm f}$ as a function of $a$, as well as the theoretical prediction (black line).
    }
    \label{fig:fig1end}
\end{figure}

\emph{Methods---}%
The experiments were performed according to the ASTM E647-15 fatigue crack growth standard~\cite{astmStandard}, where the crack length~$a$ is defined, starting from the middle of the two orifices, as can be seen from Fig.~\ref{fig:fig1end}a.
The loading frequency was $f = 1$~Hz and the cycle contrast parameter~$R$ was varied between 0.1 and 1, with the $R = 1$ case corresponding to creep (see Fig.~\ref{fig:fig1}a).
The samples were also pre-cracked to $a \approx 3$~mm according to the standard (to $a=1$-2~mm in the two-step experiments).
The material used was polymethyl methacrylate~(PMMA) in the compact tension~(CT) geometry (see~Fig.~\ref{fig:fig1end}a) defined by the standard~\cite{astmStandard}, in which one expects a priori to see a fracture process zone~(FPZ) of size~$\ell_{\rm FPZ}$~(Fig.~\ref{fig:fig1end}a).

For crack line determination the sample was imaged and the crack line position determined according to the method detailed in Ref.~\cite{lomakin2021fatigue}.
The SIF~$K$ is then determined from the crack length according to the standard~\cite{astmStandard}.

See End Matter for Detailed experimental methods.\\

Using Eqs.~\ref{eq:arrhenius} and \ref{eq:energyBarrier} one can determine the crack velocity in creep.
Extending this approach to fatigue involves taking into account the cyclic driving by integrating the instantaneous velocity given by Eq.~\ref{eq:arrhenius} over the loading cycle~\cite{savolainen2023velocity}. This gives the crack advancement per cycle
\begin{equation} \label{eq:fatigueVelocity}
    \frac{\mathrm{d} a}{\mathrm{d} N} = v_{\rm c} \int_0^{\frac{1}{f}} e^{- \beta \Delta E\left[G (t)\right]} \mathrm{d} t
\end{equation}
which can be compared with the creep velocity by computing an effective crack velocity $v = f \frac{\mathrm{d} a}{\mathrm{d} N}$.
In our case of $f=1$~Hz the effective velocities are interchangeable with $\frac{\mathrm{d} a}{\mathrm{d} N}$ (as e.g. in Fig.~\ref{fig:fig1}).

To account for the fatigue history effects we introduce an additional lengthscale contribution to the crack length~$a_{\rm f}$~(see Fig.~\ref{fig:fig1end}a),
which captures the plasticity effects which lead to the higher than expected crack velocity, i.e. the apparent additional driving of the crack line.
The integral in Eq.~\ref{eq:fatigueVelocity} can then be computed numerically using
\begin{equation}
    \Delta E = \frac{1}{\beta_{\rm d}} \left( \left[ \frac{K(a+a_{\rm f})}{K_{\rm c}} \right]^{-2 \mu} - 1 \right) .
\end{equation}
Another way to interpret this is to consider an additional energy release rate $G_{\rm f}$ which needs to be added to the observed energy release rate~$G = \frac{K^2(a)}{E}$ to yield an equivalent $K^2(a + a_{\rm f}) = K^2(a) + E G_{\rm f} $. See Ref.~\cite{supplementary} for an additional commentary discussing this equivalency.\\

\emph{Why cracks are slow or fast---}%
We address this challenge by a combination of experiment and theoretical argument. We choose as the main material used in this study~PMMA with clear ductile-brittle transition~\cite{Brown1982_ductile_brittle, Yao2022_PMMA_ductile_brittle} and strain rate dependence~\cite{WU2004_PMMA_rate_dependence}. 
The exact point where this transition happens is strongly dependent on the molecular weight, temperature and even specimen geometry~\cite{ravi1988mechanics}.
PMMA has the advantage over usual metals etc.~as fatigue study materials that it is transparent to a large degree, allowing the optical observation of the crack front and the FPZ (Fig.~\ref{fig:fig1end}a-b).
The processes happening in the FPZ, ahead and around the observed crack tip (crack closure when the stress is at the minimum, and crazing in the case of PMMA~\cite{polym15061375_crazing_01, KIM2021105719_crazing_02}) influence the crack tip velocity.
Therefore the fundamental mechanical properties of PMMA are seen in the crack velocity, as well as the FPZ size and shape~\cite{Dalbe2015PRL_PMMA_pair_of_cracks}.

Next we show how the energy release rate can be modified to yield results consistent with experimental measurements of the crack velocity in creep and fatigue of a ductile material. The goal is to find a parameter encapsulating the effect of plasticity on short cracks in fatigue, thus also combining the creep and fatigue approaches.\\

\begin{figure}[t!]
    \centering
    \includegraphics[width=\columnwidth]{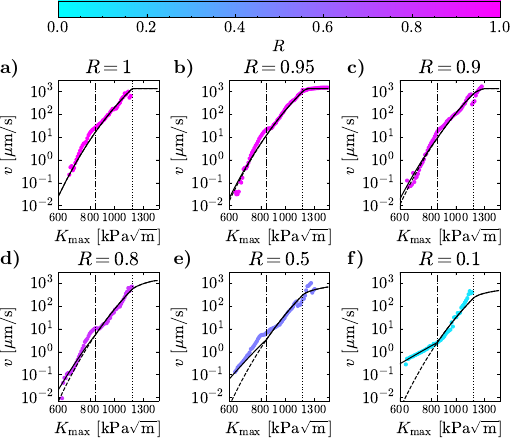}

    \caption{
    \textbf{a} to \textbf{f)} The Paris curves for different values of $R$. The solid lines are the fit, exploiting Eq.~\ref{eq:energyBarrier} with an additional lengthscale contribution and the dashed lines without this additional contribution. The vertical dashed lines indicate the value of $K_{\rm c}$, and the dashdot lines $K_0$.}
    \label{fig:paris}
\end{figure}

One can plot the Paris curves (see Fig.~\ref{fig:paris}) for each of $R$ values considered, and for the $R=1$ case (Fig.~\ref{fig:paris}a) do the creep fit utilizing (see Methods) Eqs.~\ref{eq:arrhenius} and \ref{eq:energyBarrier}. One can see that the data is fitted quite well with $v_{\rm c} = 1300$~$\mu$m/s, $\beta/\beta_{\rm d} = 9.38$, $K_{\rm c} = 1175$~kPa$\sqrt{\mathrm{m}}$, and $\mu = 0.57$. Plotting the prediction for fatigue cases (using Eq.~\ref{eq:fatigueVelocity}, see Methods) based on these values for the other $R$ values (see dashed lines in Fig.~\ref{fig:paris}) produces results that deviate from the experimentally measured values at low SIFs. The discrepancy increases with decreasing $R$.

\begin{figure}[tb!]
    \centering
    \includegraphics[width=\columnwidth]{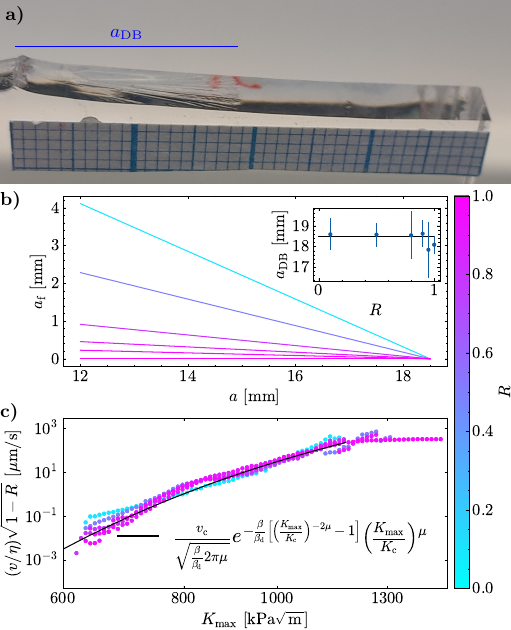}
    \caption{
    \textbf{a)}~A post-mortem image of a sample surface where we see a clear change in the crack surface morphology at a certain crack length, which we define as the ductile-brittle transition length $a_{\rm DB}$.
    \textbf{b)}~The behavior of the additional lengthscale $a_{\rm f}$ as a function of the measured crack length $a$ for different values of $R$. The inset shows the measured values of $a_{\rm DB}$ as a function of $R$ and the $a_0$ fitted to the Paris curves (black line).
    \textbf{c)}~The collapse of the curves in Fig.~\ref{fig:paris} to the creep solution when the velocities are scaled by $\eta/\sqrt{1-R}$.
    See Ref.~\cite{supplementary} for a full derivation of the creep solution and~$\eta$.
    }
    \label{fig:fig3}
\end{figure}

The crack fronts are curved (Fig.~\ref{fig:fig1end}a and also Fig.~\ref{fig:fig3}a), and in the presence of the open boundaries the shape of the front has been shown to be a parabola~\cite{barraquand2022steady} for the Edwards--Wilkinson equation on an interval (as the limiting form of the Kardar--Parisi--Zhang equation). We do not here consider the local avalanche dynamics and the front roughness nor how they couple with the interface velocity. The material properties under time-dependent loads enter as we see a clear transition point on the fracture surfaces at $a_{\rm DB}$ implying a ductile-brittle transition (Fig.~\ref{fig:fig3}a).
After this point (at high SIFs) the Paris curves follow the predictions made using the energy release rate without an additional lengthscale~(Fig.~\ref{fig:paris}).\\

A way to modify the energy release rate is to consider the additional lengthscale~$a_{\rm f}$ that needs to be added to the measured crack length~$a$ to yield the correct crack velocity.
One can see that a simple constant lengthscale~$a_{\rm f}$ is not enough to account for the increasing discrepancy with the experimental data with decreasing~$R$ but $a_{\rm f}$ needs to depend on $a$ (Fig.~\ref{fig:fig3}b). The discrepancy also vanishes above a specific SIF value~$K_0$ or specific crack length~$a_0$. This motivates the ansatz for $a < a_0$
\begin{equation} \label{eq:af}
    a_{\rm f} = a_{\rm f}^0 (1-R) \left(1 - \frac{a}{a_0} \right)
\end{equation}
(and $a_{\rm f} = 0$ for $a > a_0$).
One can see that this term is a perturbation to the creep 
limit $R \to 1$, and that the barrier-reducing effect is zero above $a_0$.
Fitting this expression to the data of Fig.~\ref{fig:paris} yields good agreement with $a_{\rm f}^0 = 13$~mm and $a_0 = 18.5$~mm ($K_0 = 843$~kPa$\sqrt{\mathrm{m}}$). 
The value of~$a_0$ is seen (inset of Fig.~\ref{fig:fig3}b) to match very well with the lengthscale~$a_{\rm DB}$ observed on the fracture surfaces. Additionally the~$a_{\rm f}$ has a natural interpretation as the FPZ size~$\ell_{\rm FPZ}$ of Fig.~\ref{fig:fig1end}a (see Ref.~\cite{supplementary} and especially Fig.~S2 for FPZ size measurements).
From the imaging data, we have extracted the FPZ size for six different points during an experiment (see Fig.~\ref{fig:fig1end}b).
Our results show that this lengthscale decreases during the experiment.
This decrease corresponds almost exactly to our theoretical treatment (Eq.~\ref{eq:af}) as can be seen from the inset of Fig.~\ref{fig:fig1end}b.
We must stress that this information has not been used in determining the fitting parameters of Eq.~\ref{eq:af}, which was done solely based on the Paris curves~(Fig.~\ref{fig:fig3}).
Using Eq.~\ref{eq:af} (see Ref.~\cite{supplementary} for details), the data of Fig.~\ref{fig:paris} then collapses in Fig.~\ref{fig:fig3}c.

We note that this lengthscale differs from the cohesive zone~(CZ) models both in behavior and in scale.
In the Dugdale model~\cite{doll1983size} the CZ size $\ell_{\rm CZ} = (\pi/8) (K / \sigma_{\rm y})^2$ (where $\sigma_{\rm y}$ is the yield stress) increases with increasing SIF, the opposite of what we observe. With $\sigma_{\rm y} = 100 \pm 40$~MPa~\cite{WU2004_PMMA_rate_dependence, mulliken2006mechanics} in the range of strain rates we have used, the CZ size is also at least an order of magnitude smaller than~$a_{\rm f}$. The CZ models might be applicable at the very beginning of an experiment before the onset of the Paris--Erdo\v{g}an regime.\\

\begin{figure}[tb!]
    \centering
    \includegraphics[width=\columnwidth]{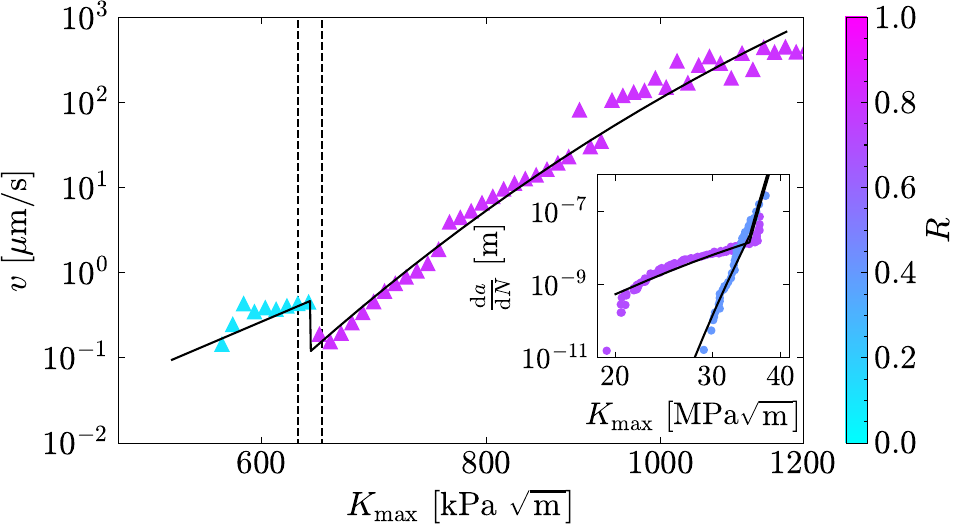}
    \caption{
    Two-step experiments with initially $R=0.10$ which is later switched to $R=0.80$ at a point corresponding to the region highlighted by dashed lines. The solid black line corresponds to the theoretical curve arising from the fit shown in Fig.\ref{fig:paris}.
    The inset shows the two curves of the steel data from Fig.~\ref{fig:fig1}b and the fit from our theory (black lines).}
    \label{fig:fig4}
\end{figure}

\emph{Predicting velocities---}%
The analysis of experimental data by the theory presented covers the range of cases where the protocol does not change, that is $R$ stays constant whether creep or fatigue, but this leaves the general case or aspect of variable loads open. Stochastic fatigue loading is in fact the typical real-life situation. Furthermore, our reasoning implies the monotonic change of $a_{\rm f}$ with crack growth and thus it is important to consider what happens if loads are varied. To this end we present an example where we challenge this approach by starting experiments with an $R$-value leading to faster growth due to the relaxation effects, and then changing it to a value close to the creep limit  $R=1$. We may derive from growth and velocity data (presented here above) suitable SIF or apparent crack length values, at which the loading or $R$ is changed so that the theory is tested. Fig.~\ref{fig:fig4} demonstrates that as expected the crack growth velocity initially decreases in a quick transient, and then starts to follow the $R=0.8$ branch. Paying now attention to the previous Fig.~\ref{fig:fig3} this means that the effective dynamical lengthscale $a_{\rm f}$ also changes to its expected value for a given nominal cracklength with the changed $R$. This would imply that exploiting Eq.~\ref{eq:af} one could incorporate varying $R$-histories and predict the crack growth.

Extending our theory to the slow-to-faster crack transition observed in the steel data of Fig.~\ref{fig:fig1}b 
involves mapping the data to an equivalent CT specimen (with $0.2 \leq a/W \leq 0.6$, see End Matter for specimen geometry details).
Using $\mu=0.57$, $\beta/\beta_{\rm d} = 40.8$, and a linearly decreasing $a_{\rm f}$ from Eq.~\ref{eq:af} yields the good fits shown in the inset of Fig.~\ref{fig:fig4}.
Due to differences in experimental geometry and lack of experimental details in the literature the exact added lengthscale value cannot be accurately determined but it starts out at around 10~\% of the crack length for the $R=0.4$ case.\\

\emph{Conclusions---}%
It has been a long-standing challenge to extend our understanding of the crack dynamics in materials to time-dependent scenarios even though this would be the main goal of statistical physics derived fracture theories. What we show here is that the complexity coming from crack closure in fatigue and a FPZ lengthscale can be harnessed into a description of the crack velocity by activated dynamics. The essence of this is an experimentally measurable length $a_{\rm f}$, that changes, or indeed lowers the energy barrier of crack dynamics. Its dependence on $R$ and $a$ shows how the crack closure effects lead to an effective FPZ size, $a_{\rm f}$ and that the effect this encodes---of how the energy release rate depends on the history and crack length $a$---decreases with crack advancement and finally vanishes at $a_{\rm DB}$.

This advancement opens up usual but challenging questions about universality and applicability. Our particular choice of material leads to the crucial effective critical exponent $\mu$ having a particular value: why this value and is this universal across materials from PMMA to metals? 
The $\mu = 0.57$ we observe corresponds to intermediate range elasticity.
In materials where the direct observation of the~FPZ is not possible the size of~$a_{\rm f}$ could be estimated by performing a creep test, fitting the model with the unmodified energy release rate, and then seeing how much faster than expected the crack grows with some~$R < 1$.
In the inset of Fig.~\ref{fig:fig4} we show how this may be done.
This needs detailed reporting of the experimental protocols~\cite{xu2024need}, not only the $\Delta K$ and $\frac{\mathrm{d} a}{\mathrm{d} N}$.

Given that our results imply more complex histories and loading scenarios can be understood and predicted, one should strive for progress into this direction as would be indeed making the connection to elastic-plastic fracture mechanics clear. An issue we have not addressed in this work is the local dynamics of crack propagation by avalanches. This is connected to the "depinning" features of the problem as the exponent $\mu$ indicates but also is an interesting problem on its own.\\

\begin{acknowledgments}
\emph{Acknowledgments---}%
M.J.A. acknowledges support from the European Union Horizon 2020 research and innovation programme under grant agreement No 857470 and from European Regional Development Fund via Foundation for Polish Science International Research Agenda PLUS programme grant No MAB PLUS/2018/8.
M.J.A. acknowledges support from the Academy of Finland (Center of Excellence program, 278367 and 317464) and the Finnish Cultural Foundation.
J.K. acknowledges the funding from Academy of Finland (308235) and Business Finland (211715).
I.V.L. acknowledges the funding from Academy of Finland (341440 and 346603).
T.M., L.T., J.L., J.K., and M.J.A. acknowledge the support from FinnCERES flagship (grant no. 151830423), Business Finland (grant nos. 211835, 211909, and 211989), and Future Makers programs.
The authors acknowledge the computational resources provided by the Aalto University School of Science ``Science-IT'' project.
\end{acknowledgments}

%

\section{End matter}

\emph{Detailed experimental methods---}%
The fatigue crack growth tests were performed according to the ASTM E647-15 fatigue crack growth standard~\cite{astmStandard}.
Compact tension~(CT) specimens were laser-cut from 5 mm poly(methyl methacrylate) (PMMA) sheets.
The final geometry of the specimens was derived based on the distance $W$ between the force application axis and the free edge of the specimen, as defined by the standard.

Sinusoidal loading with a frequency of $f = 1$~Hz and a maximum force of $F_{\rm max} = 140$~N was applied to the specimens. The cycle contrast parameter $R$, defined as the ratio of the minimum to maximum stress intensity factor ($R = K_{\rm min}/K_{\rm max}$), was varied between~0.1 and~1. The case $R = 1$ corresponds to creep. 
The samples were pre-cracked in accordance with the standard to create a natural initial notch through cyclic loading. This pre-cracking was performed using $F_{\rm max} = 140$~N and $R = 0.5$. The pre-cracking procedure was completed once the notch reached an approximate length of $a = 3$~mm. This value is approximate because the crack length measurement was not integrated to the experimental setup, requiring manual observation to determine the pre-crack length.

In experiments where $R$ was varied, the pre-crack lengths were shorter ($1-2$~mm) to enable observation of the evolving characteristic length scale $a_f$. The $R$ value was adjusted at approximately $a = 3$~mm. Again, this value is approximate due to the reasons outlined above, as also indicated by the region shown in Fig.~\ref{fig:fig4}.

Crack line tracking was performed by imaging the specimens at a frequency of 1.6 Hz using a Canon EOS~R digital camera equipped with a Canon MP-E 65~mm macro lens. To capture the crack line for subsequent tracking, the transparency of the PMMA samples was utilized. A light panel was positioned behind the specimen, while the camera was tilted~$30^\circ$ vertically relative to the specimen plane. In this configuration, the crack surface appeared dark, enhancing contrast.

An image processing algorithm was implemented, which accounted for parallax effects introduced by the optical capture system. As a result, the evolution of the crack line position as a function of time (or cycle number) was obtained. The mean crack position, $\langle a \rangle$, was used as $a$. The stress intensity factor (SIF) was then calculated according to the standard \cite{astmStandard}, using
\begin{equation} \label{eq:sif}
    K = \frac{F}{B \sqrt{W}} \frac{\left(2 + \frac{a}{W}\right)}{\left(1 - \frac{a}{W}\right)^{3/2}} \sum_{k=0}^4 p_k \left(\frac{a}{W}\right)^k
\end{equation}
where $B$ is the thickness of the sample, $W$ the width (as defined in the standard), and $p_0 = 0.886$, $p_1 = 4.64$, $p_2 = -13.32$, $p_3 = 14.72$, and $p_4 = -5.6$.

For the determination of the FPZ size~(Fig.~\ref{fig:fig1end}) we take the raw image and compute averages of the image intensity~$I$ along vertical lines for each horizontal coordinate value~$x$ (see Ref.~\cite{supplementary} and Fig.~S2 for details). 
To reduce noise, this procedure is repeated for 100~consecutive images, and the result is an average over those.
One can then take the far-field intensity and make a linear (almost constant) fit. 
The FPZ size is then the point where the measured average intensity starts to deviate from this line.

All Paris curves shown in the manuscript are averaged over several experiments (see Tbl.~\ref{tab:exps}).

\begin{table}[b!]
    \centering
    \caption{\textbf{Experimental dataset.} Used $R$ values and the corresponding number of experiments for each type of experiment shown in the manuscript.}
    \begin{tabular}{c|c}
        $R$ & number of experiments \\
        \hline
        0.1 & 4 \\
        0.5 & 5 \\
        0.8 & 5 \\
        0.9 & 5 \\
        0.95 & 7 \\
        1.0 & 9 \\
        0.1~$\to$~0.8 & 5
    \end{tabular}
    \label{tab:exps}
\end{table}

\end{document}



\title{Supplemental Material: Crack propagation by activated avalanches during creep and fatigue from elastic interface theory}

\author{Tero M\"{a}kinen}
\email{Corresponding author: tero.j.makinen@aalto.fi}
\affiliation{Department of Applied Physics, Aalto University, P.O. Box 15600, 00076 Aalto, Espoo, Finland}
\author{Lumi Tuokkola}
\affiliation{Department of Applied Physics, Aalto University, P.O. Box 15600, 00076 Aalto, Espoo, Finland}
\author{Joonas Lahikainen}
\affiliation{Department of Applied Physics, Aalto University, P.O. Box 15600, 00076 Aalto, Espoo, Finland}
\author{Ivan V. Lomakin}
\affiliation{Department of Applied Physics, Aalto University, P.O. Box 15600, 00076 Aalto, Espoo, Finland}
\author{Juha Koivisto}
\affiliation{Department of Applied Physics, Aalto University, P.O. Box 15600, 00076 Aalto, Espoo, Finland}
\author{Mikko J. Alava}
\affiliation{Department of Applied Physics, Aalto University, P.O. Box 15600, 00076 Aalto, Espoo, Finland}
\affiliation{NOMATEN Centre of Excellence, National Centre for Nuclear Research, A. Soltana 7, 05-400  Otwock-Świerk, Poland}

\date{\today}

\begin{abstract}
This is the Supplemental Material containing detailed derivation of the crack velocity in creep and fatigue conditions, a section discussing the equivalency of an additional lengthscale and an additional contribution to the energy release rate, as well as a section discussing the observed fracture process zone.

\end{abstract}

\maketitle

\section{Crack velocity and the scaling form}

In the absence of the additional lengthscale $a_{\rm f}$, the crack advancement per cycle computed from Eq.~4 
using the energy barrier given by Eq.~2 
with $G = K^2/E$ is an equation of the form $\int_0^{1/f} e^{-\lambda g^{-\mu}} \mathrm{d} t$. These can be approximated using saddle point integration~\cite{savolainen2023velocity} by taking a second order expansion $g^{-\mu} \sim g_{\rm max}^{-\mu} \left[ 1 + C \left( t - t_{\rm max} \right)^2 \right]$ around the maximum value of the integrand, corresponding to $g_{\rm max}^{-\mu}$ attained at $t=t_{\rm max}$, so that the integral becomes $e^{-\lambda g_{\rm max}^{-\mu}} \sqrt{\frac{\pi}{\lambda C} g_{\rm max}^\mu}$.

For the sinusoidally varying force
\begin{equation}
    K(t) = \frac{K_{\rm max}}{2} \left[ (1+R) + (1-R) \sin(2 \pi f t) \right]
\end{equation}
the expansion gives $C_{a_{\rm f} = 0} = 2 f^2 \mu \pi^2 (1-R)$ and thus
\begin{equation} \label{eq:fatigueVelocitySaddle}
    v(a) 
    \approx 
    \frac{v_{\rm c}}{\sqrt{\frac{\beta}{\beta_{\rm d}} 2 \pi \mu (1-R)}} 
    e^{-\frac{\beta}{\beta_{\rm d}} \left[ \left(\frac{K_{\rm max}}{K_{\rm c}}\right)^{-2 \mu} - 1 \right]}
    \left(\frac{K_{\rm max}}{K_{\rm c}}\right)^{\mu}.
\end{equation}\\

When the contribution of $a_{\rm f}$ is included, one can take the first order expansion
\begin{equation}
    v(a+a_{\rm f}) \approx v(a) + \frac{\mathrm{d} v}{\mathrm{d} a_{\rm f}} \bigg|_{a_{\rm f} = 0} a_{\rm f}
\end{equation}
where (using Eq.~\ref{eq:fatigueVelocitySaddle})
\begin{equation}
    \frac{\mathrm{d} v}{\mathrm{d} a_{\rm f}} \bigg|_{a_{\rm f} = 0}
    = \frac{\mu v}{K_{\rm max}}
    \left[
    \frac{2 \beta}{\beta_{\rm d}}
    \left( \frac{K_{\rm max}}{K_{\rm c}} \right)^{-2 \mu}
    + 1
    \right]
    \frac{K_{\rm max}}{W} Q
\end{equation}
and
\begin{equation} \label{eq:q}
    Q(a) =
    \frac{1}{2}
    \frac{8 + \frac{a}{W}}{\left( 1 - \frac{a}{W} \right) \left( 2 + \frac{a}{W} \right)}
    +
    \frac{
    \sum_{k=0}^3 (k+1) p_{k+1} \left(\frac{a}{W}\right)^{k}
    }{\sum_{k=0}^4 p_k \left(\frac{a}{W}\right)^k}
\end{equation}
where the coefficients $p_k$ are the ones used in the definition of the SIF (Eq.~7
).\\

To collapse the Paris curves one then simply needs to compute the quotient of the crack velocities with and without the additional contribution
\begin{equation}
    \eta = \frac{v(a+a_{\rm f})}{v(a)} \approx
    1 + \mu
    \left[
    \frac{2 \beta}{\beta_{\rm d}}
    \left( \frac{K_{\rm max}(a)}{K_{\rm c}} \right)^{-2 \mu}
    + 1
    \right]
    Q(a) \frac{a_{\rm f}}{W}
    .
\end{equation}
Substituting Eq.~6 
to this finally gives
\begin{equation}
    \eta = 1 + \mu \frac{a_{\rm f}^0}{W}
    \left[
    \frac{2 \beta}{\beta_{\rm d}}
    \left( \frac{K_{\rm max}(a)}{K_{\rm c}} \right)^{-2 \mu}
    + 1
    \right]
    Q(a) (1-R) \left( 1 - \frac{a}{a_0} \right)
\end{equation}
indicating that $v/\eta$ should collapse to Eq.~\ref{eq:fatigueVelocitySaddle} for all fatigue curves ($R > 0$). Additionally, multiplying by $\sqrt{1-R}$ gets rid of the $R$~dependence and produces the collapse seen in Fig.~4c
.

\section{Equivalency of lengthscale and energy release rate}

In the theoretical treatment of the main text
we have added the additional contribution (described by Eq.~6
) as an additional lengthscale.
Another possibility would be to consider an additional contribution to the energy release rate by setting $G = K^2(a)/E + G_{\rm f}$.
The form of Eq.~6 
can be very closely approximated by replacing $K^2(a+a_{\rm f})$ by $K^2(a) + E G_{\rm f}$ where
\begin{equation} \label{eq:Gf}
    G_{\rm f} = G_{\rm f}^0 (1-R) \left(1 - \frac{K}{K_0} \right)
\end{equation}
where again $G_{\rm f} = 0$ for $K > K_0$. The corresponding constants are $E G_{\rm f}^0 = 800$~MPa$\sqrt{\mathrm{m}}$ and $K_0 = 843$~kPa$\sqrt{\mathrm{m}}$.\\

To see the equivalency one can start by taking the first order expansion of $K^2$ with respect to the additional lengthscale
\begin{equation} \label{eq:add_len_k2}
    K^2(a+a_{\rm f}) \approx K^2(a) + \frac{2K^2(a)}{W} Q(a) a_{\rm f}
\end{equation}
where $Q$ is again given by Eq.~\ref{eq:q}. If the additional lengthscale~$a_{\rm f}$ and additional energy release rate~$G_{\rm f}$ are equivalent
\begin{equation} \label{eq:add_en_k2}
    K^2(a+a_{\rm f}) = K^2(a) + E G_{\rm f} .
\end{equation}

\begin{figure}[tb!]
    \centering
    \includegraphics[width=\columnwidth]{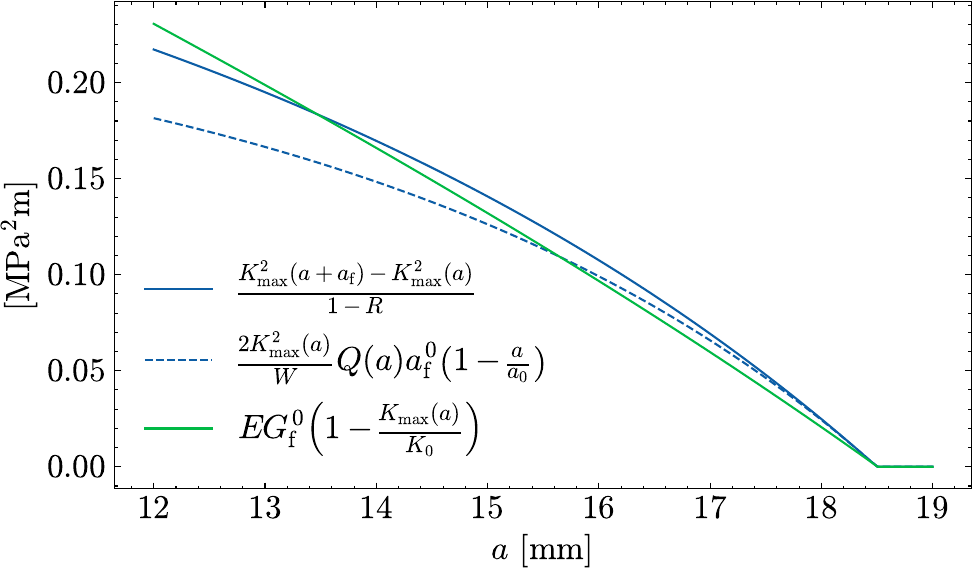}
    \caption{\textbf{The equivalency of additional lengthscale and additional energy relase rate.} The additional contributions to the square of the SIF for the additional lengthscale (blue) and the additional energy release rate (green).}
    \label{fig:fig5}
\end{figure}

The additional contributions to the square of the maximum~SIF, with the~$(1-R)$ terms removed, are plotted in Fig.~\ref{fig:fig5}. The solid blue line corresponds to the true addition with the additional lengthscale, blue dashed line to the approximation of Eq.~\ref{eq:add_len_k2} (with Eq.~6 
substituted),
and the green line to the addition of energy release rate (Eq.~\ref{eq:Gf} substituted into Eq.~\ref{eq:add_en_k2}). One can clearly see that the nonlinearity of the additional contribution due to an additional lengthscale is small, so the linear contribution from an additional energy release rate matches this well.
This can be understood by realizing that $K^2$ is approximately linear in $a$ (a result which is exact in
a straight crack perpendicular to the loading direction in an infinite plane of uniform stress).

\section{Observed fracture process zone}

\begin{figure}[tb!]
    \centering
    \includegraphics[width=\columnwidth]{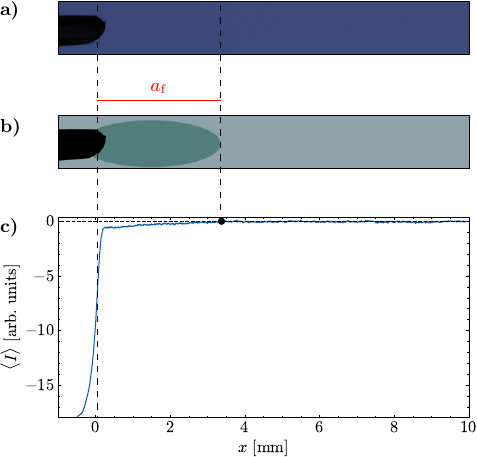}
    \caption{The lengthscale measurement in PMMA is done using the optical variation of the material in the FPZ and outside. The raw image (panel a) separates to the open crack, the FPZ and the region outside the FPZ (schematic in panel b). The FPZ region cannot be seen by naked eye but can be seen in the average image intensity $\langle I \rangle$ in front of the crack, the black dot marking the end of this region (panel c).
    }
    \label{fig:sfig3}
\end{figure}

In Fig.~5 of the main text we show that the history effect and the additional lengthscale are (at least mostly) recoverable, i.e. only visible within the experiment. 
The previous or several previous cycles define the size of~$a_{\rm f}$.
The crack velocity corresponds to the current $R$-value, seen in the quick transient in the crack velocity at the $R$-value change. This is well encompassed by the theoretical prediction of the size of $a_{\rm f}$ (based on Eq.~6 of the main text) which drops from $a_{\rm f}(R=0.1) \approx 4$~mm to $a_{\rm f}(R=0.8) \approx 1$~mm at the protocol change. 
This indicates that the type of deformation or "damage" in the region of additional lengthscale $a_{\rm f}$ is recoverable.\\

Figure~\ref{fig:sfig3} shows the relation of this additional lengthscale~$a_{\rm f}$ to the raw data (Fig.~\ref{fig:sfig3}a).
It shows a simple image (or light) intensity analysis from the beginning of an experiment with $R=0.1$ where the largest $a_{\rm f}$ (see schematic in Fig.~\ref{fig:sfig3}b) is observed. 
Microcracking or crazing is usually seen as completely black areas as the mirrorlike surfaces create a "dark field" reflecting the light away from the camera lens. 
Thus the crack front is seen completely black. 
PMMA in particular has a strong crazing effect followed by craze initiated voids growing perpendicular to the applied stress~\cite{polym15061375_crazing_01, KIM2021105719_crazing_02} making it simple to follow the crack growth.
Also this is seen in Fig.~\ref{fig:sfig3}c, where there is a sharp drop around the average crack front location at $x=0$~mm (corresponding to~$a$).
We can also see a more subtle change in the intensity of the light traveled through the sample lasting up to 4~mm from the crack tip, corresponging to our additional lengthscale~$a_{\rm f}$.
This effect is due to the photoelasticity where the intensity of the light passing through the sample is reduced due to the stress induced polarization~\cite{Dalbe2015PRL_PMMA_pair_of_cracks, CHRISTOPHER2008_photoelasticity}.